\documentclass[preprint]{iucr}

\usepackage[version=4]{mhchem}
     \papertype{FA}               
     \journalcode{S}              

\begin{document}                  

\title{The Heisenberg-RIXS instrument at the European XFEL}

\author[a,*]{Justine}{Schlappa}
\aff[a]{European XFEL, Holzkoppel 4, Schenefeld, 22869, \country{Germany}}

\author[b,c,\#]{Giacomo}{Ghiringhelli}
\aff[b]{Dipartimento di Fisica, Politecnico di Milano, piazza Leonardo da Vinci 32, I-20133 Milano, \country{Italy}}
\aff[c]{CNR-SPIN, Dipartimento di Fisica, Politecnico di Milano, I-20133 Milano, Italy}

\author[a]{Benjamin E.}{Van Kuiken}
\author[a]{Martin}{Teichmann}
\author[a]{Piter S.}{Miedema}
\author[a]{Jan Torben}{Delitz}
\author[a]{Natalia}{Gerasimova}
\author[a]{Serguei}{Molodtsov}
\author[a]{Luigi}{Adriano}
\author[a]{Bernard}{Baranasic} 
\author[a]{Carsten}{Broers} 
\author[a]{Robert}{Carley}
\author[a]{Patrick}{Gessler}
\author[a]{Nahid}{Ghodrati}
\author[a]{David}{Hickin}
\author[a]{Le Phuong}{Hoang}
\author[a]{Manuel}{Izquierdo}
\author[a]{Laurent}{Mercadier}
\author[a]{Giuseppe}{Mercurio}
\author[a]{Sergii}{Parchenko}
\author[a]{Marijan}{Stupar}
\author[a]{Zhong}{Yin}

\author[b]{Leonardo}{Martinelli}
\author[a,b]{Giacomo}{Merzoni}
\author[b,d]{Ying Ying}{Peng}
\aff[d]{Present address: International Center for Quantum Materials, School of Physics, Peking University, Beijing 100871, \country{China}}

\author[e]{Torben}{Reuss}
\author[e,f]{Sreeju}{Sreekantan Nair Lalithambika}
\author[e,f,\dagger]{Simone}{Techert}
\author[e,g,\circ]{Tim}{Laarmann}
\aff[e]{Deutsches Elektronen-Synchrotron DESY, Notkestraße 85, 22607 Hamburg, \country{Germany}}
\aff[f]{Institute of X-ray Physics, Goettingen University, Friedrich Hund Platz 1, 37077 Goettingen, \country{Germany}}
\aff[g]{The Hamburg Centre for Ultrafast Imaging CUI, Luruper Chaussee 149, 22761 Hamburg,
\country{Germany}}

\author[h]{Simo}{Huotari}
\aff[h]{Department of Physics, University of Helsinki, P.O. Box 64, FI-00014 Helsinki, \country{Finland}}

\author[i]{Christian}{Schroeter}
\author[i]{Burkhard}{Langer}
\author[i]{Tatjana}{Giessel}
\aff[i]{BESTEC GmbH, Am Studio 2b, 12489 Berlin, \country{Germany}}

\author[j]{Robby}{Buechner}
\author[j]{Jana}{Buchheim}
\author[j]{Vinicius}{Vaz da Cruz}
\author[j]{Sebastian}{Eckert}
\author[j]{Grzegorz}{Gwalt}
\author[j,k]{Chun-Yu}{Liu}
\author[j]{Frank}{Siewert}
\author[j]{Christian}{Sohrt}
\author[j]{Christian}{Weniger}
\author[j]{Annette}{Pietzsch}
\author[k,l]{Stefan}{Neppl}
\author[k]{Friedmar}{Senf}

\author[a,\%]{Andreas}{Scherz}
\author[j,k,\$]{Alexander}{F\"ohlisch}

\aff[j]{Institute Methods and Instrumentation for Synchrotron Radiation Research, Helmholtz-Zentrum Berlin für Materialien und Energie GmbH, Albert-Einstein-Straße 15, 12489 Berlin, \country{Germany}}
\aff[k]{University of Potsdam, Institute of Physics and Astronomy, Karl-Liebknecht-Straße 24/25, 14476 Potsdam, \country{Germany}}
\aff[l]{Present address: Paul Scherrer Institut, Forschungsstrasse 111, 5232 Villigen PSI, \country{Switzerland}}

\aff[*]{justine.schlappa@xfel.eu}
\aff[\#]{giacomo.ghiringhelli@polimi.it}
\aff[\%] {andreas.scherz@xfel.eu}
\aff[\dagger] {simone.techert@desy.de}
\aff[\circ] {tim.laarmann@desy.de}
\aff[\$]{alexander.foehlisch@helmholtz-berlin.de}


\keyword{resonant inelastic X-ray scattering}\keyword{RIXS}\keyword{hRIXS}\keyword{Heisenberg RIXS}\keyword{X-ray Raman scattering}\keyword{free electron laser}\keyword{FEL}\keyword{XFEL}\keyword{European XFEL}\keyword{soft X-ray}\keyword{time-resolved spectroscopy}\keyword{spin dynamics}\keyword{photochemistry}\keyword{resonant X-ray diffraction}\keyword{X-ray resonant diffraction}\keyword{XRD}\keyword{charge order}\keyword{spin order}

\maketitle                        

\begin{synopsis}
The aim of this work is to present Heisenberg RIXS (hRIXS) at the European XFEL, the first high-resolution soft X-ray instrument enabling time-resolved resonant inelastic X-ray scattering close to the transform limit from solid and liquid-jet samples. The optical design, mechanical layout and performance are described in detail and discussed in broader context of X-ray spectroscopy at high-repetition Free-Electron Lasers.
\end{synopsis}

\begin{abstract}
\emph{Abstract} Resonant Inelastic X-ray Scattering (RIXS) is an ideal X-ray spectroscopy method to push the combination of energy and time resolutions to the Fourier transform ultimate limit, because it is unaffected by the core-hole lifetime energy broadening. And in pump-probe experiments the interaction time is made very short by the same core-hole lifetime. RIXS is very photon hungry so it takes great advantage from high repetition rate pulsed X-ray sources like the European XFEL. The hRIXS instrument is designed for RIXS experiments in the soft X-ray range with energy resolution approaching the Fourier and the Heisenberg limits. It is based on a spherical grating with variable line spacing (VLS) and a position-sensitive 2D detector. Initially, two gratings are installed to adequately cover the whole photon energy range. With optimized spot size on the sample and small pixel detector the energy resolution can be better than 40\,meV at any photon energy below 1000\,eV. At the SCS instrument of the European XFEL the spectrometer can be easily positioned thanks to air-pads on a high-quality floor, allowing the scattering angle to be continuously adjusted over the 65-145\,$^{\circ}$ range. It can be coupled to two different sample interaction chamber, one for liquid jets and one for solids, each equipped at the state-of-the-art and compatible for optical laser pumping in collinear geometry. The measured performances, in terms of energy resolution and count rate on the detector, closely match design expectations. hRIXS is open to public users since the summer of 2022. 
\end{abstract}


\section{Introduction}

At the core of the quantum mechanical description of matter stand uncertainty relations as identified by Heisenberg in 1927 \cite{Heisenberg1927}. The position-momentum uncertainty results from the commutation of the position and momentum operators linked to the Planck constant $\hbar$. In contrast, time-energy uncertainty $\delta E \delta t \geq \hbar$ does not stand on such fundamental footing, 
reflecting the fact that time is a variable. How to conceptualize time and time-energy uncertainty relations has been debated over many years \cite{Heisenberg1927,MandelstamTamm1945,AhaBohm1961} and led to numerous reviews \cite{Busch2008,Hilgevoord2005}. The value of $\hbar = 0.66$\,eV\,fs 
makes the time-energy uncertainty central for physics, chemistry, biology, and materials science. It relates the femtosecond (fs) timescale of reaction dynamics and materials functionality to the electron-Volt (eV) energy scale of valence electrons, chemical bonds as well as spin and magnetic properties.

 Unlike for an isolated quantum-mechanical system, the lower limit for the uncertainty relation during a (disturbing) measurement process is given as $\delta E \delta t \geq 2\pi\,\hbar$ 
 \cite{Messiah1999}. The trade-off between energy and time has important implications for spectroscopic and time-resolved techniques in the study of matter. In particular, a parameter that is hard to access in the energy (frequency) domain can be more easily determined in the time domain and vice versa. Time-resolved spectroscopy has thus been rapidly developing in the optical regime thanks to ultra-short laser pulses, which are used both to initially bring the system out of equilibrium (pump) and to observe the transient modifications of its properties (probe) - encoded in the complex optical constants. More recently, the pump-probe scheme has been extended to photoelectron spectroscopy, X-ray absorption and X-ray scattering, which provide direct insights on the electronic structure and ordering phenomena. The idea is that (by refining both the pump and the probe) it will become possible to recognize links among the various microscopic degrees of freedom (atomic positions, charge density, spin orientation) by selectively exciting one of them and observing changes induced in the other ones. 

The advent of X-ray free electron lasers (XFELs) kicked off the age of selective pump-probe experiments based on X-ray photons. Initially, the inherent intensity and energy fluctuations of the Self-Amplified Spontaneous Emission (SASE) process that creates ultrashort X-ray laser pulses implied that the power density could easily exceed the radiation damage threshold for a majority of materials. This fact effectively limited the average flux actually usable, making photon hungry techniques challenging at low repetition rate XFELs. That is the reason why inelastic X-ray scattering in the time-resolved mode has taken off more slowly than techniques with higher signal at the detector, i.e. diffraction, coherent scattering and X-ray absorption spectroscopy. These limitations are now overcome thanks to high repetition rate XFELs based on superconducting linac technology, where the energy per X-ray pulse can be adjusted to avoid damaging the sample while preserving the number of photons hitting the sample per second at the level of storage ring-based experiments, or even higher. Resonant inelastic X-ray scattering (RIXS) appeared to be the best technique to perform high quality time-resolved spectroscopy at high repetition rate XFELs for a number of reasons. Firstly, it does not involve charged particles (i.e. electrons) that are affected by space charge. Secondly, time-resolved RIXS is sensitive to all low-energy excitations of charge, spin, orbital polarization as well as structural distortions and their ultrafast dynamics \cite{ament2011resonant,gelmuk2021}. Therefore, RIXS is the ideal spectroscopic tool to advance the science of how to govern materials properties, control chemical and biological processes, or create novel and transient phases that cannot be reached in equilibrium. In particular, since RIXS is resonant Raman scattering in the X-ray regime, both Stokes and anti-Stokes features can be used. The latter being univocal markers of the excited states, they can be used to characterize the transient states and their dynamics with superior elemental and chemical selectivity and stringent symmetry selection rules. As a photon‐in/photon-out technique RIXS can access all aggregate states in equilibrium and non-equilibrium. Strong external stimuli such as laser pulses and electromagnetic fields allow for state preparation but do not affect the measurement. Beyond that, the full potential of novel non‐linear processes (e.g. four-wave mixing with X‐rays) can be explored.

The choice of RIXS entails considering another time scale that is defined by the intrinsic lifetime of the intermediate state involved in the second-order resonant scattering process. The Heisenberg relation holds here too, but with little impact on the ultimate time resolution of the pump-probe RIXS experiment, because RIXS usually involves core holes with lifetime of few fs at most, much shorter than the experimental pump-probe resolution. The core-hole lifetime determines the apparent duration of the resonant scattering process, because within the intrinsic energy width of intermediate states, multi-path interference and energy-detuning effects are significant. These crucial mechanisms are well captured by the Kramers-Heisenberg equation \cite{gelmuk2021}. The intermediate state lifetime is a different timescale, independent of the transient state prepared by the pump pulse. As long as the latter are much longer than the former, the RIXS process can be regarded as instantaneous. Conversely, if and when the experimental resolution will reach the femtosecond range and the inelastic X-ray scattering will be used to study truly ultrafast dynamics, some precautions will be needed to avoid misinterpretation of the experimental results \cite{Hilgevoord1996,Hilgevoord1998}. 

At the European X-ray Free Electron Facility GmbH (European XFEL), which provides unprecedented ultra‐short soft X‐ray pulses with high repetition rate and brightness, the conceptual aspects outlined above can be put into experimental reality and we can finally push time-resolved X-ray spectroscopy closer to the limits of retrievable information. We call this approach Heisenberg RIXS (hRIXS). 
In this article, we describe the hRIXS spectrometer for soft X-ray photons (200-2000\,eV), which is coupled with two experimental environments, one for solid (named XRD chamber) and one for liquids samples (CHEM chamber).

The Heisenberg RIXS approach, proposed by a User Consortium and endorsed by the European XFEL management but largely supported by external funding [Supplementary Materials], led to the design and construction of a high-resolution soft X-ray spectrometer to be installed at the SCS instrument.   
Below we describe the technical design of the hRIXS spectrometer and its performance space. We discuss the underlying criteria and constraints, the final optical design and the actual realization, including the mechanical and detection aspect. The target performances were quickly reached, as confirmed in the commissioning runs of 2021 and 2022. An outlook to all operational parameters is given at the end of the article. 

\section{Requirements and goals of a RIXS experimental setup operating at an X-ray FEL}

\subsection{Soft X-ray RIXS spectroscopy}

RIXS is an energy loss spectroscopy performed with X-ray photons with and energy tuned to the binding energy of a core level of one of the atomic species present in the material. The resonance greatly enhances the scattering cross section and provides chemical and site selectivity. Moreover, the spin-orbit interaction in the intermediate state core level is often large enough ($>5$\,eV) to trigger pure spin-flip transitions, allowing the study of magnons with RIXS \cite{ament2011resonant}. In contrast to inelastic neutron scattering (INS) RIXS enables also observation of non-spin-flip magnetic excitations \cite{Schlappa2018}. The sizable momentum of X-rays allows the study of collective excitations or quasi-particles in both, energy and momentum domain (already in the soft X-ray range). The potential of RIXS has emerged in the last 15 years, after the experimental bandwidth has been improved enough to resolve the physically relevant local and collective excitations in the samples. The bandwidth limit to resolve these excitations is around 100-120\,meV, although 40\,meV is a standard value in the best facilities and 20\,meV is technically feasible in some cases. The task is technically challenging, because the RIXS experiment requires very high resolving power both in the monochromator preparing the beam before the sample and in the spectrometer analyzing the scattered radiation: 100\,meV bandwidth at 1000\,eV photon energy requires 15000 resolving power on each instrument. The low efficiency of the scattering process and the high resolving power imply that a brilliant undulator source, optimized beamline optics to monochromatize the beam and refocus it on a few-micron spot size on the sample, and an efficient spectrometer are all needed to measure high-resolution RIXS spectra.

\begin{figure}
    \centering
    \includegraphics[width=0.7 \textwidth]{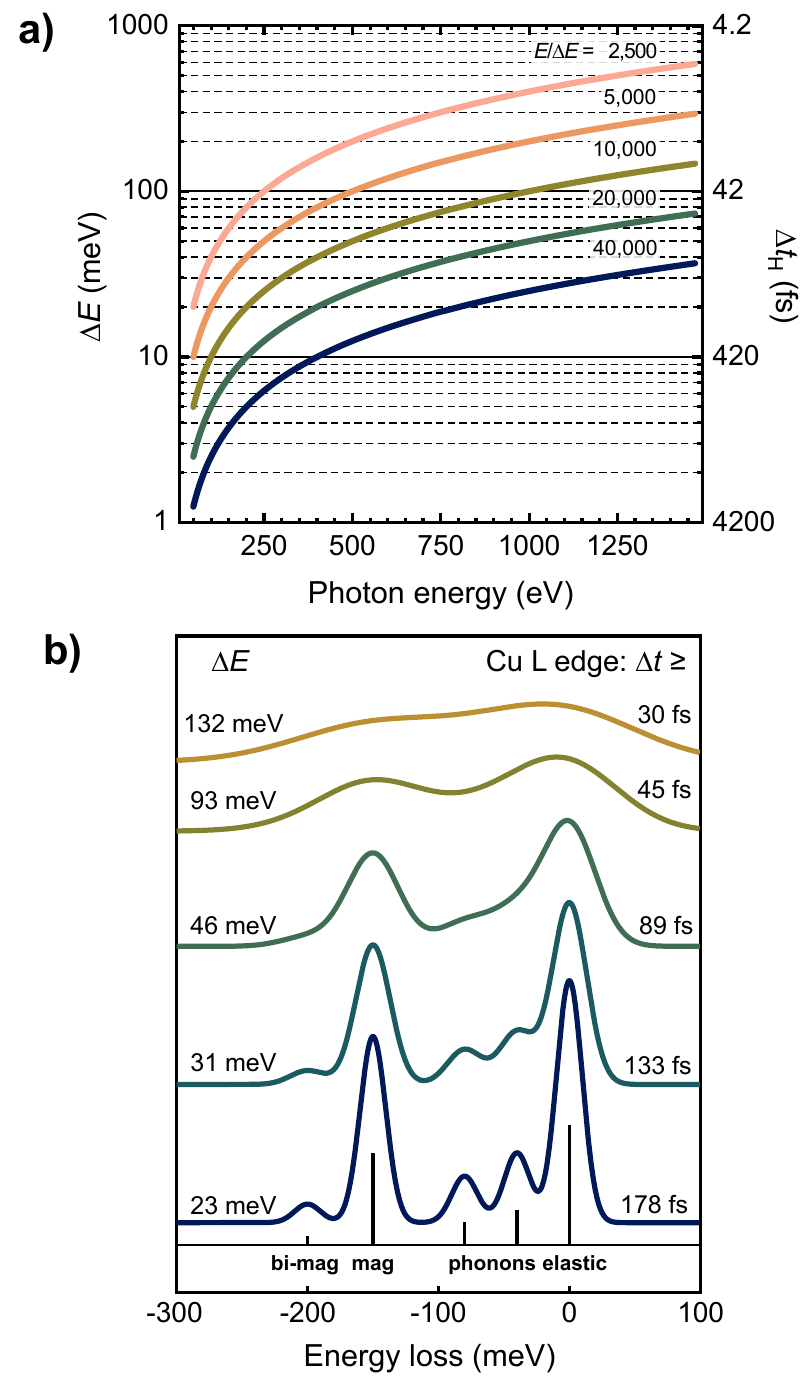}
    \caption{Visualization of Heisenberg uncertainty for time and energy during a soft X-ray spectroscopy experiment. Lower limits for energy and time resolution at given energy resolving power (a). We notice that resolution of 30\,meV (present resolution limit for RIXS at synchrotrons at 1000\,eV photon energy) sets the time resolution limit to about 138\,fs (that is within the scope of SCS \cite{Gerasimova2022mono} and other European XFEL instruments). Modelling low-energy RIXS spectra for a correlated copper-oxide material\cite{ament2011resonant}, depending on the instrumental energy resolution (b).}
    \label{fig:Heisenberglimit}
 \end{figure}

The reason for choosing the soft X-ray range for high-resolution RIXS is motivated by the presence of the K absorption edges of light elements such as, C, N and O, and the L$_{2,3}$ edges of $3d$ transition metals in this energy range, broadly used also in X-ray absorption spectroscopy. Furthermore, oxygen and $3d$ transition metals are the key ingredients of a huge number of materials with intriguing electronic and magnetic properties. Probably the most essential examples for the solid state physics community are cuprate superconductors, layered Cu-O compounds showing superconductivity well above the liquid nitrogen boiling point at 77\,K. That explains why RIXS has been developed mostly in the soft X-ray range. Moreover, RIXS can be used to study molecules, either in gas, liquid, or solid state \cite{hennies2010resonant,yin2015ionic,cheng2022, kotani2001}. In particular, understanding the oxygen and nitrogen electronic structures is essential for a huge number of substances, whereas organometallic compounds are of particular interest when they contain a $3d$ transition metal. 

\subsection{Soft X-ray RIXS instrumentation}
Due to the low count rate, the first RIXS instruments were based on compact spectrometers, with wide angular acceptance, spherical gratings with constant line spacing mounted in Rowland geometry and microchannel-plate detectors at very grazing incidence \cite{nordgren1986design,nordgren1989soft,callcott1986}. The use of spherical gratings with variable line spacing (VLS) allowed to mount detectors at less grazing incidence \cite{Osborn1995VLS,dallera1996soft} and to reduce the detector motion range, thus allowing to design longer spectrometers with intrinsic higher resolving power. With the advent of CCD detectors at affordable price, good quantum efficiency and sufficiently small pixel sizes, the soft X-ray RIXS eventually reached resolving power better than 2000 in the beginning of the years 2000 \cite{dinardo2007gaining}, and entered the high-resolution age ($E/dE \ge 10000$) in 2007, thanks to the first combined design of the ADRESS beamline and the 5\,m long SAXES spectrometer at the PSI/SLS \cite{ghiringhelli2006saxes,strocov2010high}. A few years later larger (8-12\,m long) and more ambitious instruments were built at the TPS \cite{lai2014highly}, ESRF\cite{brookes2018beamline}, DLS\cite{zhou2022i21}, and NSLS II \cite{dvorak2016towards} or under construction at NanoTerasu \cite{miyawaki2022design} and Sirius \cite{rodrigues2019}. In most of the cases the spectrometer optical layout is based on VLS spherical gratings, coupled to parabolic collecting mirrors to increase the angular acceptance in the non-dispersive direction.

The hRIXS spectrometer was designed to be the first high-resolution spectrometer at a free-electron laser. The goal was to guarantee a maximum flexibility to fulfill the various expectations of the different users. Therefore, it had to allow a very high-energy resolution, matched to that of the beamline, possibly getting close to the Heisenberg limit allowed by the time structure of the FEL. Alternatively, a different configuration based on single-pulse detection and using ultrafast detectors has to be present to push the time resolution at the expenses of relaxed energy resolution and larger spot size at the samples. Here, it is interesting to take a look at the theoretical limits of time and energy resolution given by the uncertainty relations, and express them in relation to the resolving power of the instrumentation. A summary of this time-energy landscape and influence on RIXS spectra is presented in Figure \ref{fig:Heisenberglimit}.
Moreover, hRIXS has to allow studying multi-photon RIXS-excitations by using a high photon-flux density of the focal spot on the sample. Finally, the facility must allow the continuous change of the scattering angle, and has to be easy to setup and operable with different experimental stations and sample environments.

\subsection{SCS beamline characteristics}
  
The SCS instrument is one of three soft X-ray beamlines at the European XFEL, located at SASE3 \cite{tschentscher2017}. The photon energy of the SCS instrument covers the soft and tender X-ray regime, from 280\,eV to 3000\,eV. The European XFEL, operating in the self-amplified spontaneous emission (SASE) regime, produces intense fs pulses up to MHz repetition rate with very high degree of transverse coherence (close to 100\%) but limited longitudinal one. Within the typical pulse duration of few fs to tens of fs the coherence time is mostly in the sub-fs range, resulting in tens or hundreds of longitudinal modes. The SASE3 grating monochromator is meant to substantially reduce the intrinsic bandwidth (0.3-1\% in the soft X-ray range) and to improve the longitudinal coherence. Achieving close to transform-limited pulses after the monochromator is a major challenge, limited by the quality of the optical elements, in particular of the grating. Both the figure error and accuracy of VLS spacing increase the time$\times$bandwidth product with respect to what is expected for ideal optics. A transmission $>4 \sigma$ of the Gaussian beam profile is foreseen to approach the diffraction limit, but gratings that long and with that quality cannot be manufactured yet, so shorter gratings have been implemented. Presently, the monochromator operates with two gratings. The low-resolution grating is optimized for time-resolved experiments (few to few tens of fs RMS) and moderate resolving power (2000–5000). The high-resolution grating reaches a resolving power of 10000 at the cost of larger pulse stretching in time. The time$\times$bandwidth product of these gratings is estimated to be close to ideal below 500\,eV and increases at higher photon energies, doubling around 780\,eV and 1240\,eV for the low-resolution and high-resolution grating, respectively.  
The $10^{-4}$-$10^{-3}$ transmission results in a pulse energy ranging 0.1-10\,$\mu$J at the sample. The details of the SASE3 monochromator can be found in \cite{Gerasimova2022mono}.
The focus size at the SCS instrument is variable, controlled by Kirkpatrick-Baez optics. The bendable mirrors allow to change the vertical and horizontal focus independently from 1\,mm down to $\sim$1\,$\mu$m  \cite{Mercurio:22}. For RIXS studies a horizontally elongated beam is particularly convenient, with a small vertical size (typically $\sim 10$\,$\mu$m) and 100-500\,$\mu$m horizontally, matched to the pump laser beam size at the sample.

\section{Design of the spectrometer and its environment}

\subsection{Constraints for the design}

Several constrains had to be taken into account for the design and construction of the hRIXS spectrometer. The experimental hall of the European XFEL (where all instruments are located) is placed underground. This puts severe limitations on the available space, meaning that the total length of the hRIXS spectrometer could not exceed 5\,m. The range of possible positions of the refocusing focal point combined with the size of the experimental hutch impose the range of the accessible scattering angle $2\Theta$ to be 65\,$^{\circ}$-145\,$^{\circ}$ (an overview can be found in the supplementary material).
Moreover, the spectrometer cannot be permanently installed, because the SCS instrument hosts a range of interchangeable setups dedicated to different experiments. Therefore, the hRIXS spectrometer has been designed to be retractable while at the same time minimizing the time for putting it on- and off-line, and, even more importantly, to properly align it and start measuring, so to make the best use of the beamtime. In addition, the hRIXS spectrometer has to be coupled to different interaction chambers and sample environments, e.g. a goniometer mounting for solid single crystals or a liquid jet setup for molecular systems.

The European XFEL operates in burst mode at 10\,Hz. Every 0.1\,s a burst of pulses is emitted during a time window of 600\,$\mu$s. Up to 400\,$\mu$s of that time can be dedicated to SASE3, that would result in a train of pulses separated by a minimum interval of 885\,ns. Therefore, one long pulse train can deliver up to 452 FEL pulses at the rate of 1.13\,MHz to SASE3 and up to 4520\,pulses/s. Due to the SASE process, the spectral distribution and the intensity vary widely from pulse to pulse. Advanced diagnostics at European XFEL allows to measure single-shot and average pulse intensities for data normalization \cite{Maltezopoulos:xt5015}.

\begin{figure}
    \centering
    \includegraphics[width=0.8 \textwidth]{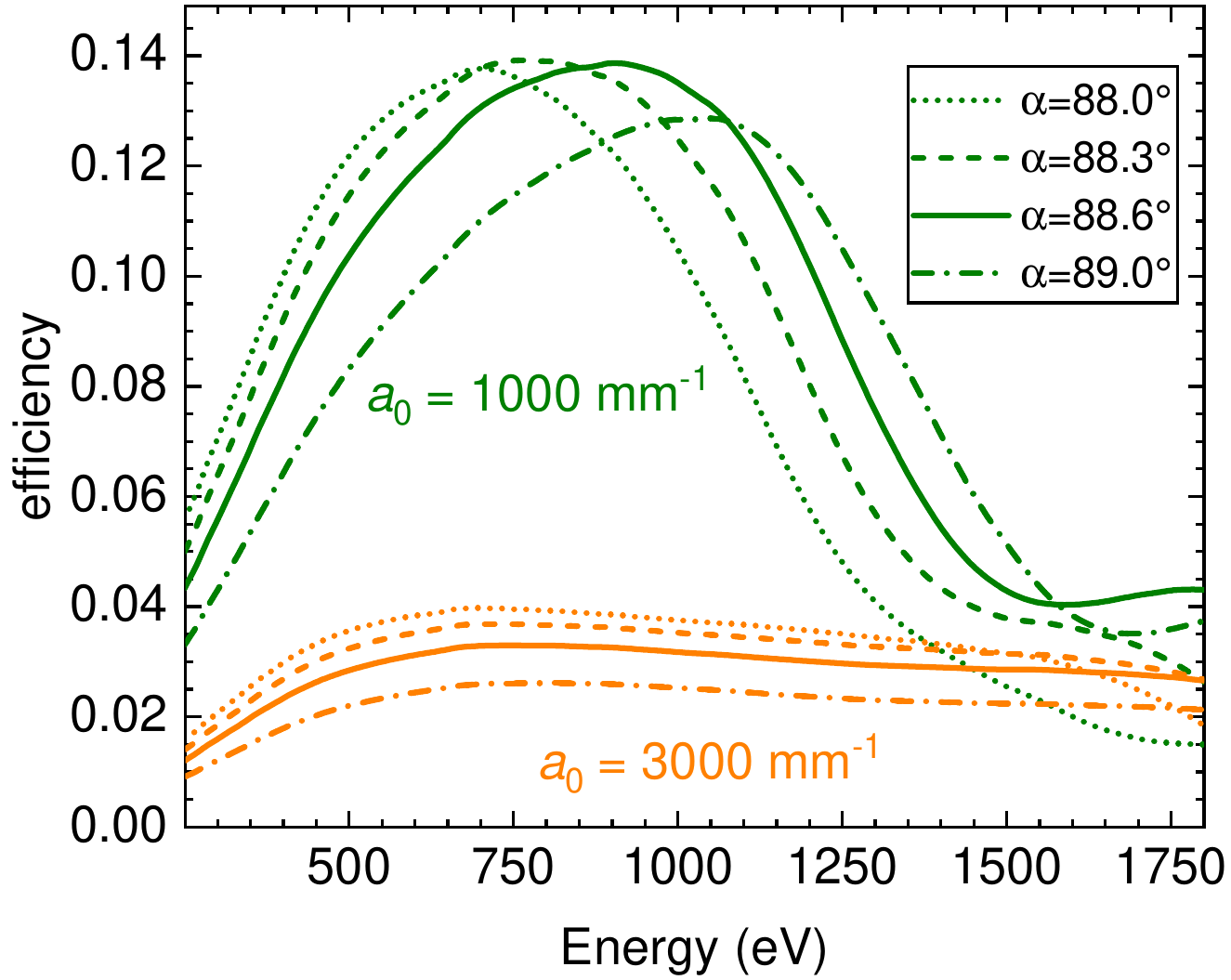}
    \caption{hRIXS spectrometer computed grating efficiency at 4 incidence angles with Au coating, aspect ratio $c/d=0.60$ (0.65) and groove depth $h=5$\,nm (9\,nm) for high-resolution grating (HRG) and high-transmission grating (HTG).}
    \label{fig:GratEffic}
 \end{figure}

This effective repetition rate is extremely demanding for detectors, particularly when combined with high spatial resolution, large area, and large number of pixels as for RIXS spectrometers. There are three options for operation, which are given by the readout time of the detector:

\begin{itemize}
\item pulse-resolved detection

\item train-resolved detection

\item integrating detection
\end{itemize}

Pulse-resolved detection requires high readout time (1\,$\mu$s range or better). In addition, the large amount of data from a two-dimensional detector requires data-storage capabilities, as the data cannot be send out at that rate. Pulse-resolved detection would be the preferred choice, giving the possibility to normalize the data shot-by-shot, to take reference data in parallel, and to allow achieving the highest temporal resolution. However, at present there are only few detectors that operate at that rate and they have moderate spatial resolution. Examples are based on microchannel-plate photon converter and delay-line current pulse detection with position encoding, as well as mega-pixel size detectors built of Si-sensors \cite{Porro2021}. Train-resolved detection requires moderate reading and transfer speed (better than 100\,ms). The temporal resolution is limited by the jitter within each train (that is already smaller than jitter between single trains and considerably smaller than FEL pulse stretching due to the monochromator). Examples of detectors that can currently work in that mode are based on electron-amplified CCD and CMOS technology. Any detector that requires acquisition times of 1\,s or longer has to be operated in integration mode. The "standard" high-resolution detectors, CCDs, have to be operated in that mode.

\subsection{Optical layout}
We have worked to find the best compromise in terms of throughput, resolution and operative flexibility given the diverse and often contrasting needs of the different operation modes foreseen for the hRIXS instrument. For the study of low-energy excitations in solid samples the highest resolving power is the priority, whereas in other cases the temporal resolution represents the limiting factor and energy resolution can be compromised. The former has implications on the choice of detector (fast detectors capable of resolving single X-ray pulses have poor spatial resolution with respect to the CCD/CMOS detectors that necessarily integrate over thousands of X-ray bunches). Moreover, the beam spot size on the sample has to be tunable to adopt the X-ray beam size to the sample size (e.g. for different liquid jets) and the focal size of the pump laser that can vary strongly with wavelength and required pulse energies. This allows to find a trade-off for possibly high signal level, while staying below the damage threshold of particularly solid samples by X-ray and pump laser beams. The overall  X-ray intensity can be decreased by tuning a gas attenuator located in the SASE3 tunnel.

\begin{figure}
    \centering
    \includegraphics[width=0.95 \textwidth]{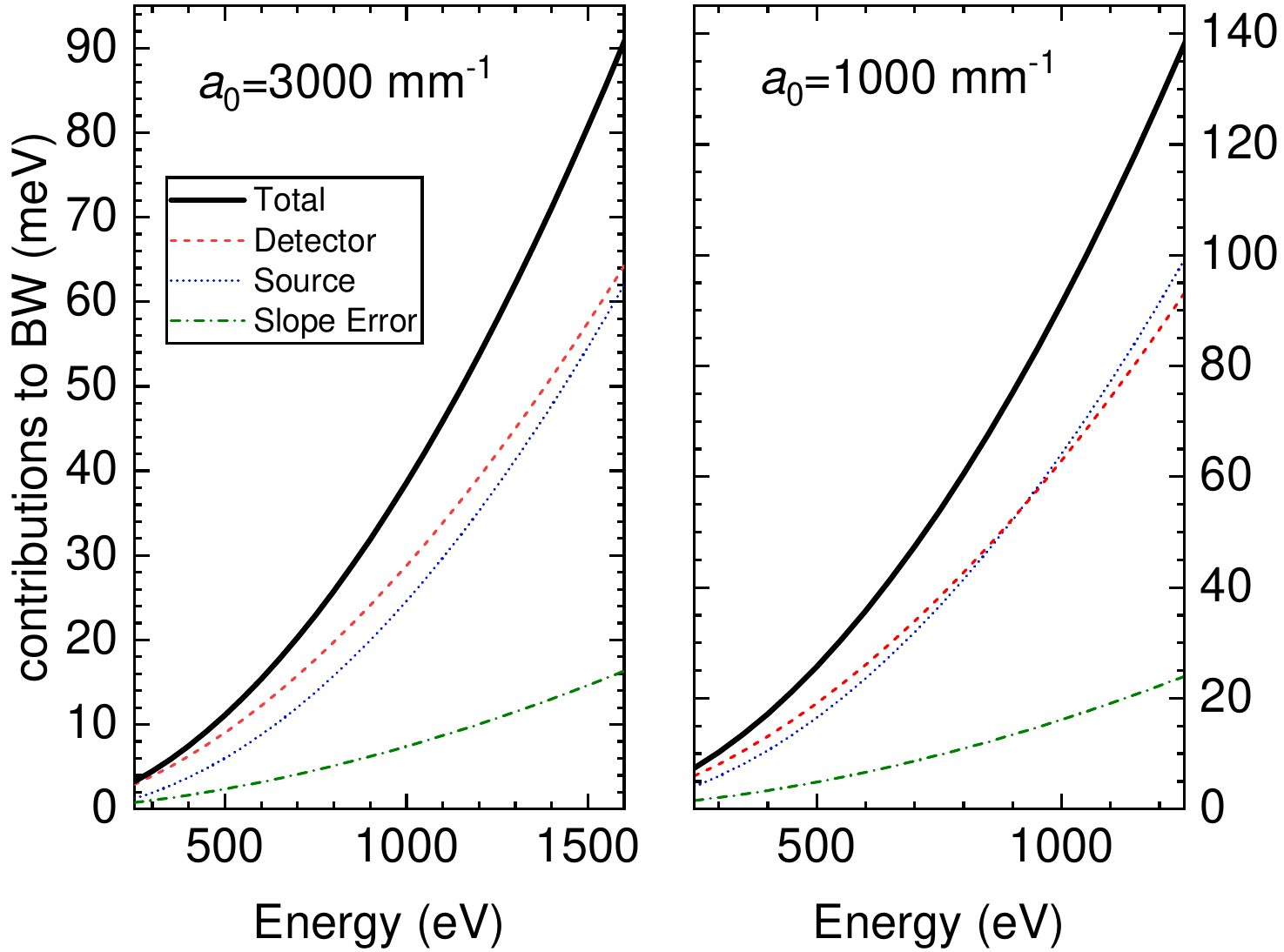}
    \caption{Contributions to the spectrometer energy bandwidth for the two gratings: The total resolution is given by the quadratic combination of the partial values. Calculations made for the nominal values of source size and detector spatial resolution used for the optimization of the VLS parameters ($S_1=5$\,$\mu$m, $S_2=10$\,$\mu$m), and slope error $s'=0.1$\,$\mu$rad\,rms.}
  \label{fig:ResolutionContrib}
 \end{figure}

We have initially explored the performances of a single VLS spherical-grating spectrometer over an extremely wide range of parameters. This elementary optical layout has several advantages for an instrument of about 5\,m of total length. The total range of positioning of the detector is easily manageable with standard mechanical actuators and the whole instrument can be held on a single girder, while keeping the internal alignment. The mechanical stability is provided by the high-quality floor of the experimental hutch. A single or double parabolic mirror usually added in longer spectrometers (ERIXS at ID32/ESRF \cite{brookes2018beamline}, I20 at DLS \cite{zhou2022i21}, SIX at NSLS II \cite{dinardo2007gaining}) would bring a relatively small gain in luminosity, at the cost of relevant mechanical additional complications in the small available space and would not work properly with a horizontally wide beam to be adopted in most of the experiments (the parabolic mirrors cannot work properly with a mm-sized source). Therefore, having fixed the maximum length (sample to detector distance) given by the physical space available in the SCS experimental hutch, we have over a wide range explored the 4 key parameters of the spectrometer: 
\begin{enumerate}
\item the vertical spot size on the sample surface, $S_1$

\item the actual spatial resolution for the 2D position-sensitive detector, parallel to the detector surface, $S_2$

\item the central grating ruling density, $a_0$

\item the entrance arm length, $r_1$
\end{enumerate}

The other parameters (grating radius of curvature $R$; VLS polynomial coefficients of the ruling density $a_1$, $a_2$, $a_3$; angle of incidence on the grating $\alpha$; focusing conditions, with cancellation of coma aberration, $r_2$ and $\beta$) are calculated following the procedure used by Ghiringhelli et al. \cite{ghiringhelli2006saxes}. The (grazing) angle of incidence on the detector $\gamma$ is fixed; for the calculations a reference value is used, typically around 20$^\circ$, for which the effective spatial resolution of the detector is improved by a factor $(\sin{\gamma})^{-1} \simeq 4.4$ and the quantum efficiency of Si-based detectors (CCD or CMOS) is still close to 1. The analytical calculations were then cross-checked and refined with Shadow ray tracing \cite{sanchez2011shadow3} and with the code by Strocov for spherical VLS-based spectrometer design \cite{strocov2011numerical}. 
The boundary conditions were set by the maximum total length ($L=5000$\,mm) and by the minimum $r_1 = 900$\,mm. Two gratings ($a_0 = 3000$\,mm$^{-1}$ and $a_0 = 1000$\,mm$^{-1}$) are sufficient to cover the whole working energy range 250-1600\,eV with good flexibility. After a preliminary comparison of the performances of spectrometers optimized on five different combinations of source size and detector effective spatial resolution we have selected $S_1 =5$\,$\mu$m and $S_2 =10$\,$\mu$m for the final optimization. The choice of the two groove densities was dictated by the need of maximizing dispersion, while keeping a decent efficiency ($a_0=3000$\,mm$^{-1}$, 1st order diffraction efficiency better than $\eta > 3\%$) in one case, and to maximize efficiency with a reasonable resolving power ($a_0=1000$\,mm$^{-1}$, $\eta > 12\%$) in the other case (see Figure\,\ref{fig:GratEffic}). \\

\begin{table}
\caption{The main parameters for the high-resolution grating (HRG) and the high-transmission grating (HTG) of the hRIXS spectrometer.} 
\begin{tabular}{lcc}      
     & HRG         & HTG        \\
\hline \hline
$E_0$ (mm)      & 900      & 600      \\
$R$ (mm)      & 64647      & 70114      \\
 $a_0$ (mm$^{-1})$     & 3000      & 1000      \\
  $a_1$ (mm$^{-2})$     & 1.232      & 0.446      \\
  $a_2$ (mm$^{-3}$)     & $5.35 \times 10^{-4}$      & $2.06 \times 10^{-4}$       \\
  $a_3$ (mm$^{-4})$     & $3.5 \times 10^{-7}$      & $1.6 \times 10^{-7}$        \\
 actual $s'$ ($\mu$rad\,rms)     & 0.18      & 0.11          \\
area (mm $\times$ mm)     & 200 $\times$ 40      & 200 $\times$ 40       \\
$c/d$      & 0.60      & 0.65       \\
 $h$ (nm)      & 5      & 9          \\
 micro-roughness (nm)    & 0.2      & 0.5          \\
 coating thickness (nm)    & 29      & 40          \\
 \\
\end{tabular}
\label{tab:GratingParameters}
\end{table}

As shown in Figure\,\ref{fig:ResolutionContrib}, this design has many excellent features. The resolving power is higher than 30000 (20000) for photon energies below 1000\,eV (500\,eV) using the high-resolution grating - HRG with $a_0=3000$\,mm$^{-1}$ (high-transmission grating - HTG with $a_0=3000$\,mm$^{-1}$). The optimization of performances requires a good optical matching between the beamline and the spectrometer, and is obtained with a minimum of optical components and an entrance-arm length that varies little with energy. The resolving power obviously decreases with larger spot size and worse detector resolution. A slope error $s' = 0.1$\,$\mu$rad rms for the grating has been considered all along the calculations. This value is extremely good, at the limit of the present technology for spherical gratings. However, as it will become evident below, a less stringent slope error (up to 0.25\,$\mu$rad rms) would not alter significantly the resolution of the proposed optical layout. 

 \begin{figure}
    \centering
    \includegraphics[width=0.95 \textwidth]{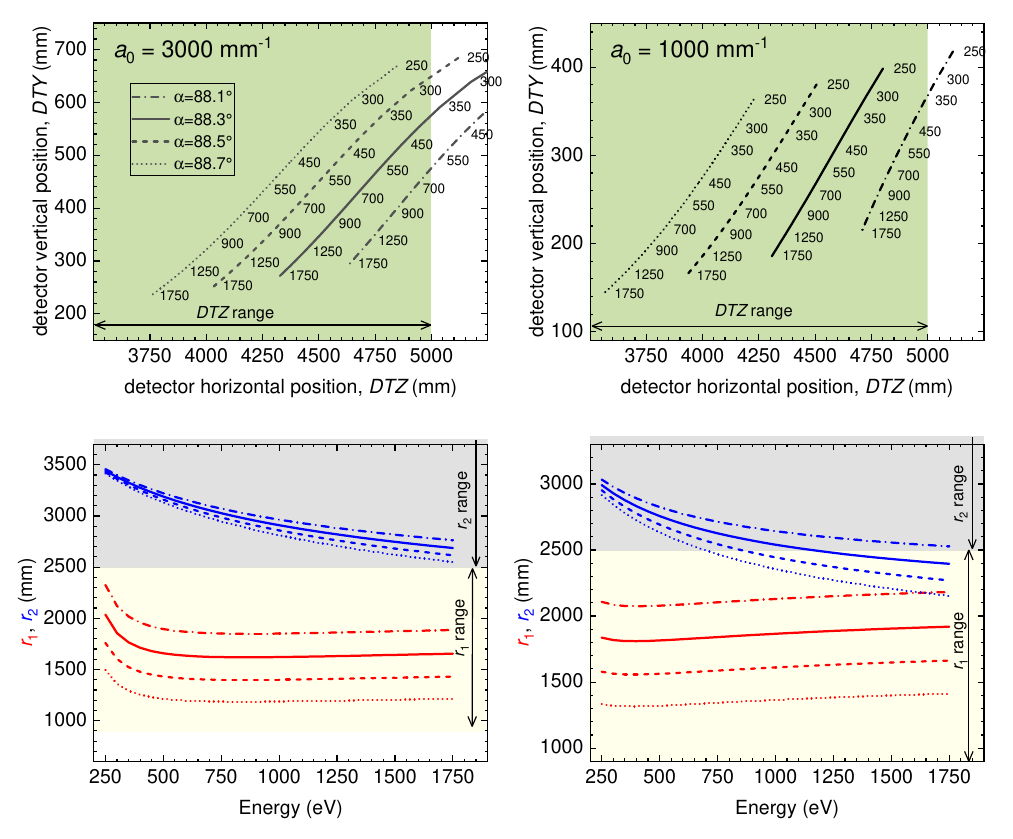}
    \caption{Working points for the two gratings, as function of the photon energy and of the incident angle $\alpha$.}
    \label{fig:positions}
 \end{figure}

The VLS parameters $R, a_1, a_2$ were optimized using the criterion that, at a reference energy $E_0$, the contributions to the total spectrometer bandwidth coming from $S_1$ and $S_2$ are equal (see Figure\,\ref{fig:ResolutionContrib}: This balance is respected over the whole energy range), having fixed the reference total length $L_0=r_1 + r_2$ and the grazing-incidence angle on the detector $\gamma=20 ^\circ$. In the VLS parameter optimization process it is important to check that the full energy range is covered while fulfilling the coma aberration minimization criterion. The eventual parameters are summarized in Table\,\ref{tab:GratingParameters}.  For all gratings the laminar trapezoidal groove profile was made on single-crystal silicon substrates with final reflective Au coating. The high-transmission grating was produced by Precision Gratings (HZB) on a substrate from Pilz (measured sagittal slope error $s' = 0.11$\,$\mu$rad\,rms). The high-resolution grating was produced by HORIBA Jobin Yvon on a substrate from Zeiss ($s' = 0.18$\,$\mu$rad\,rms). The deviation of the VLS parameters from specifications was smaller than 2\% (up to the 2nd order).

\begin{figure}
    \centering
    \includegraphics[width=0.99 \textwidth]{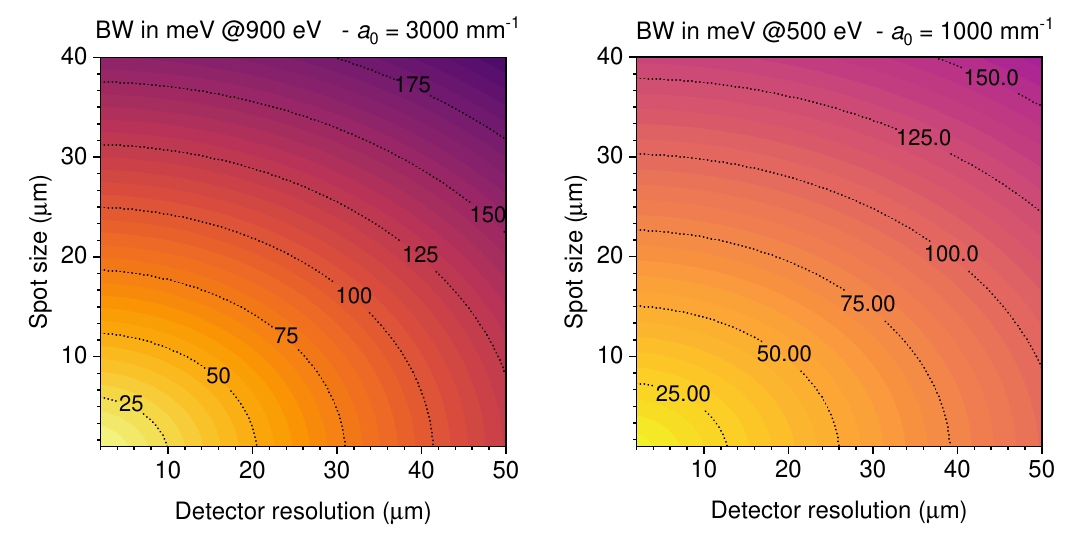}
    \caption{Spectrometer energy bandwidth for the two gratings as function of the spot size $S_1$ and of the detector spatial resolution $S_2$. The total resolution is given by the quadratic combination of the partial values. Calculations made for the nominal value of the slope error $s'=0.1$\,$\mu$rad\,rms.}
\label{fig:S1S2dependence}
\end{figure}

Both gratings can cover the 250-1750\,eV photon energy range with the appropriate choice of the $(\alpha,r_1)$ parameter combination that cancel the coma aberration, as shown in Figure\,\ref{fig:positions}. The HRG, optimized for the L$_{2,3}$ edges of $3d$ transition metals, can be used with $88.1^\circ \leq \alpha \leq 88.7^\circ$ for every energy above 600\,eV. The HTG, designed for the lower energy range, can be used with $88.2^\circ \leq \alpha \leq 88.7^\circ$ already from 250\,eV. The best choice of $(\alpha,r_1)$ will be determined case by case, optimizing the actual efficiency and resolution. Those depend also on other parameters, such as the beamline monochromator, the beam size on the sample, and the detector performances. An overview of the influence of $S_1$ and $S_2$ on the final resolution is shown in Figure\,\ref{fig:S1S2dependence}.

A third grating for the tender X-ray energy range (from 1500\,eV to 3000\,eV) is foreseen with a dedicated grating holder. Calculations indicate that resolving power better than 15000 and efficiency larger than 1.2$\%$ for $a_0=3000$\,mm$^{-1}$ and $\alpha \sim 89.0^\circ$ are possible in the whole range.

\subsection{Mechanical design}

\begin{figure}
    \centering
    \includegraphics[width=0.95 \textwidth]{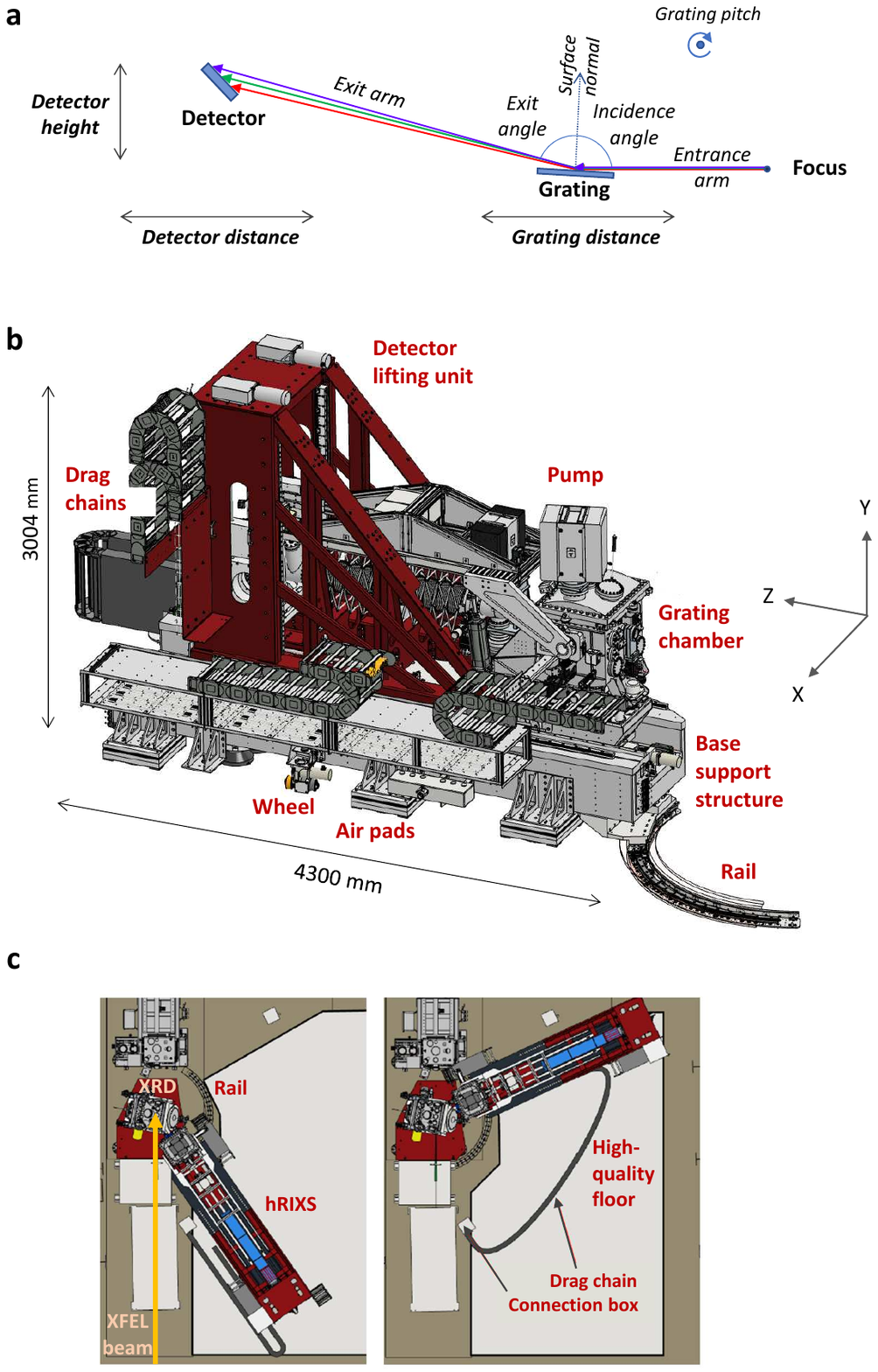}
    \caption{hRIXS spectrometer overview: diffraction scheme and degrees of motion for operation (a), model of the spectrometer (b), top view of spectrometer when placed at the interaction point at the two extreme scattering angles: back-scattering at $2\Theta=145^{\circ}$ (left) and forward-scattering at $t\Theta=65^{\circ}$ (right) (c).}
    \label{fig:hRIXSscheme}
\end{figure}

There are two optical elements inside the hRIXS spectrometer – the grating and the detector. This is the minimum number of elements required for adjustment of a spherical VLS grating spectrometer. Since the spectrometer had to be easily adjustable, a collecting mirror was not foreseen. In order to align the spectrometer for one photon energy, four degrees of freedom are required: The incidence angle on the grating $\alpha$, the entrance arm length $r_1$, the exit angle $\beta$, and the exit-arm length $r_2$. These are realized by the following four motions: grating pitch angle (changing the incidence angle $\alpha$), the horizontal translation of the grating chamber (changing the distance from the grating to the focus $r_1$), the horizontal translation of the detector chamber and the vertical translation of the detector chamber (both changing the distance between the grating and the detector $r_2$ and the exit angle $\beta$), see Figure\,\ref{fig:hRIXSscheme}a. The required movement range for the translation of the two chambers is large (around 1.5\,m), to allow operating the spectrometer over a large photon energy range with one grating. At the same time the stability requirements are extremely high (in the $\mu$m-range). The working points of the spectrometer are summarized in Figure\,\ref{fig:positions}. Table\,\ref{tab:hRIXSmotion} displays the motion range (see also Supplement for stability measurement data). \\

\begin{table}
\caption{hRIXS spectrometer motion ranges. For stability specified (measured) values are shown.}
\label{tab:hRIXSmotion}
\begin{tabular}{lll}      
\hline \hline 
Description    &Range        &Stability              \\ 
\hline 
Grating pitch, $\alpha$                     & 83$^{\circ}$ - 92$^{\circ}$         & $< 0.09 \: (0.06)\,\mu$rad       \\
(grx)                     &         &      \\
Grating horizontal        & 900\,mm - 2500\,mm        & $< 50\,\mu$m                   \\
motion, $r_1$ (gtx)        &       &  \\
Detector horizontal      & 3500\,mm - 50000\,mm      & $< 50\: (0.03)\,\mu$m      \\
motion (dtx)      &     &      \\
Detector vertical    & 1410\,mm - 2210\,mm     & $< 3\,\mu$m  \\
motion (dty)    &    &   \\
Scattering angle,   &65$^{\circ}$ - 145$^{\circ}$      &  \\
$2\Theta$ (rry)    &     &  \\
\hline
\\
\end{tabular}
\end{table}

\begin{figure}
    \centering
    \includegraphics[width=0.85 \textwidth]{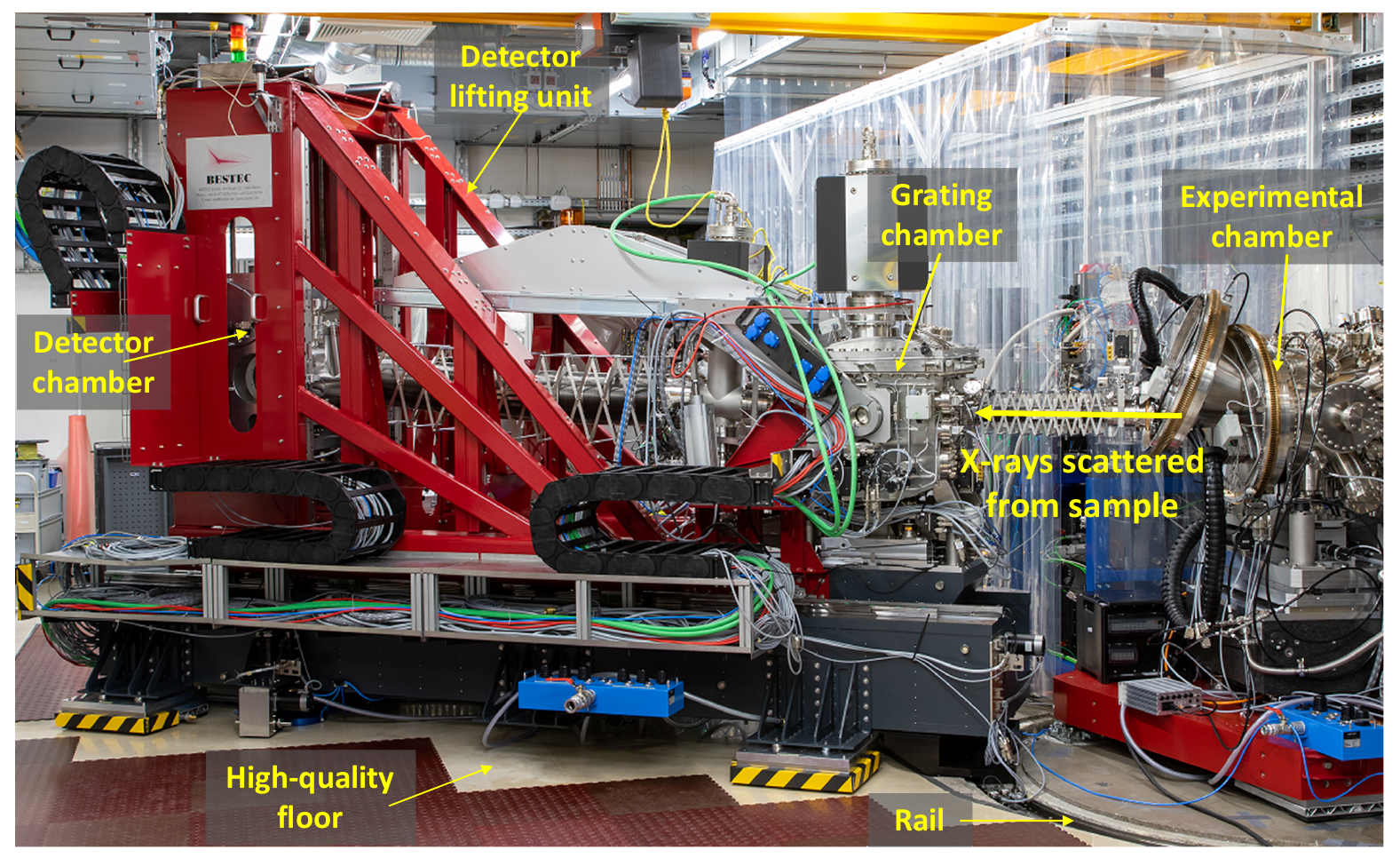}
    \caption{Photograph of the hRIXS spectrometer installed in the working point inside the SCS hutch (the high-quality floor is partially covered for protection).}
    \label{fig:hRIXSfoto}
\end{figure}

The model of the spectrometer is shown in Figure\,\ref{fig:hRIXSscheme}b. The spectrometer is placed on air-pads on top of a high-quality floor. The bottom part of the support consists of a large mineral casting base, to ensure very good thermal and mechanical stability (for more details see Supplemental Material). The grating chamber has an additional mineral casting support, which also holds the inner mechanics. The detector chamber is held by a steel lifting. 
When the spectrometer is placed at the interaction point, rails inside the experimental floor guide the rotation of the spectrometer in order to scan the scattering angle. Figure\,\ref{fig:hRIXSscheme}c shows spectrometer positions for the maximum and minimum scattering angles. When not in use, the hRIXS spectrometer can be detached and placed in a parking position, away from the interaction point. The weight of the spectrometer is approximately 10\,t and the dimensions roughly 4.2\,m x 2.1\,m x 3.1\,m (L x W x H). Figure\,\ref{fig:hRIXSfoto} shows a photograph of the hRIXS spectrometer, when connected to the interaction point.

Due to the small vertical X-ray spot size at the SCS instrument, there is no need for an entrance slit for the hRIXS spectrometer. A mask unit, placed in front of the grating, gives the possibility to narrow down grating illumination. The inner mechanics of the grating chamber provides space for three gratings. Exchange of gratings is realized by a horizontal transverse motion. There are additional degrees of motion for grating alignment. Behind the grating tank an aperture unit with four independent aperture blades can be used to shield stray light falling on the detector. The detector chamber is optimized for large detectors. The detector mounting flange has the size of 300\,DN and is tilted at $\gamma=25^{\circ}$ to the exit arm. In order to keep the incidence angle fixed at the detector and also to prevent the large bellows connecting the grating chamber with the detector chamber from breaking, the detector chamber needs to be rotated when changing the detector position. 

\subsection{Controls}
Using the Karabo control system \cite{hauf2019karabo}, the movement of the spectrometer is done in the physically relevant coordinates (i.e. entrance- and exit-arm distances and angles) instead of individual motor positions. It also assures that the spectrometer does not move outside its safe working parameters.
The movement of the scattering angle involves lifting and rotating the spectrometer, which needs to be followed precisely by the motion of the endstation in order not to break the connecting bellows. All this motion is orchestrated by the Karabo control system, while the real-time parts are written using TwinCat 3 by Beckhoff Automation.
The same control system is also used to automate standard experimental procedures, like $\theta-2\Theta$-scans, knife-edge scans or pump-probe delay scans.

\section{Sample environment} 
     
\subsection{Pump-probe laser}
The pump-probe (PP) laser operates in the same mode as the FEL (burst mode at 10\,Hz). There are currently two working points, 1.1\,MHz, which can deliver pulse energies up to 0.2\,mJ for 800\,nm pump (the fundamental) and 113\,kHz, which can deliver factor 10 higher pulse energies. Depending on future experimental needs further working points will be implemented. Other wavelengths that are available at SCS instrument are: 400\,nm (2nd harmonic generation), 266\,nm (3rd harmonic generation) and wavelenghts in the range of 2.5\,$\mu$m down to 350\,nm (that can be generated using an OPA). The minimum laser spot size is in the range of 100\,$\mu$m (depending on wavelength and the focal distance). Linearly (variable) and circularly polarized laser pump is available. There is a laser-incoupling unit placed 2\,m upstream from the interaction point that is available for nearly-collinear incoupling. Incoupling schemes at shorter focal distance are available as well. 

    \subsection{Chemical sample environment}
\begin{figure}
    \centering
    \includegraphics[width=0.95 \textwidth]{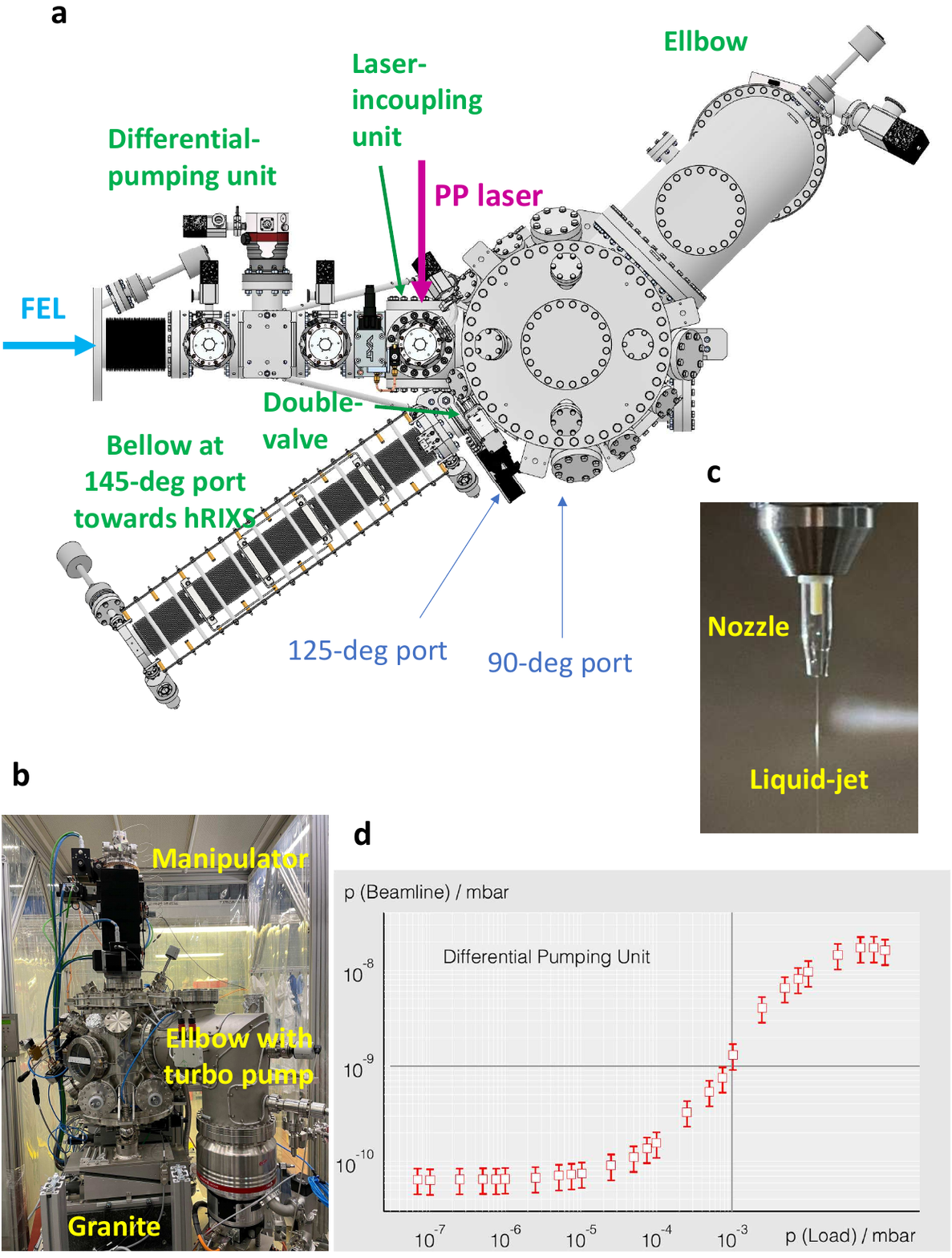}   
    \caption{CHEM setup: Model with top view (a), photograph of the CHEM chamber (b) photograph of the liquid jet (c) performance of the differential pumping unit (d).}
    \label{fig:Chem}
\end{figure}

     Time-resolved pump-probe RIXS experiments of chemical systems in their natural environment (the liquid phase) is enabled in the experimental CHEM endstation shown in Figure\,\ref{fig:Chem}. This poses several technical requirements in terms of sample-delivery system, photon-diagnostic tools and vacuum conditions. 

    The experimental main chamber is a DN300 tube with CF300 flanges on top and bottom, CF150 entrance flange towards the beamline, CF40 connections to the hRIXS spectrometer in $2\Theta=90^\circ$, 125$^\circ$ and 145$^\circ$ scattering angle with respect to the incoming FEL beam. For the installation of optics, diodes and other experiment relevant devices, a board with M6 threaded holes is mounted inside the chamber. A CF250 flange connects the pumping unit to the main chamber with a HiPace\,2300 turbomolecular pump. Two cold traps are installed for the collection of the liquid sample and improving the vacuum during operation. The whole endstation sits on an adjustable support structure with five degrees of freedom for an optimal alignment of the FEL beam through the apertures of the differential pumping stage (DPS). The sample manipulator is placed on the top CF150 flange of the CHEM vessel and consists of DDF 63 rotary feedthrough and PMM 12 XYZ-manipulator from VAb (the feedthrough is mounted above the manipulator). Table\,\ref{tab:CHEMinnerMec} shows overview of the available degrees of freedom for the sample motion.\\

\begin{table}
\caption{Motion ranges for CHEM sample environment.}
\label{tab:CHEMinnerMec}
\begin{tabular}{llll}      
\hline \hline 
 Motion    &Range       &Accuracy    &Description        \\ 
\hline 
$\theta$      &$\pm$180$^{\circ}$      &0.01$^{\circ}$    &Sample incidence angle      \\ 
$tx$      &$\pm$7.5\,mm       &1\,$\mu$m  &Translation perpendicular to FEL \\ 
$tz$     &$\pm$7.5\,mm      &1\,$\mu$m   &Translation along FEL     \\ 
$ty$      &100\,mm   &   &Vertical translation \\ 
\\
\end{tabular}
\end{table}

    The liquid sample is injected into the interaction point via the liquid-microjet technology consisting of a HPLC pump from Shimadzu and a quartz capillary nozzle with an orifice in the order of around 15\,$\mu$m till 50\,$\mu$m. A switching device allows to change between up to six different samples. 
    The custom-designed nozzle holder is part of a multipurpose unit, which additionally include tools for diagnostics at the interaction point (to check the FEL and laser beam size via the knife-edge method, fluorescence screens to optimize the spatial overlap between laser and FEL beams and a diode as well as a Si$_3$N$_4$ membrane for finding the laser-FEL time overlap). The holder is also suitable to mount solid samples for studies with the CHEM setup.

    The high vapour pressure of liquid samples and the strict vacuum conditions of the beamline require a highly efficient DPS to overcome the several orders of magnitude in vacuum differences, which is connected between the last valve of the SCS beamline and the gate valve before the laser incoupling chamber. With a running liquid jet the usual pressure in the main chamber is in the order of $10^{-3}$\,mbar and the beamline interlock is around $10^{-8}$\,mbar. To accomplish the vacuum difference, the DPS consists of three pumping units with an aperture of 2.5\,mm diameter each made from B$_4$C.  
    The DPS can overcome a pressure difference of more than 5 orders of magnitude. The vacuum of the hRIXS spectrometer is protected by a membrane that is mounted on the double-valve at the CHEM vessel facing the hRXIS connection.

    The laser incoupling unit consists of a CF100 chamber with six ports and is connected to the entrance of the main chamber and a gate valve to the DPS. It accommodates a motorized 2-inch mirror mount, which itself can be adjusted by a motorized XYZ-manipulator. An aperture with a rectangular shape shields the optics from the spraying of the liquid jet.

    \subsection{Solid sample UHV environment} 
    
An ultra-high vacuum (UHV) environment down to $10^{-9}$\,mbar is provided by the SCS instrument baseline chamber for X-ray resonant diffraction (XRD). This setup is equipped with a diffractometer for time-resolved and non-linear XRD studies. It offers a cryogenic sample environment for solid samples in the temperature range from ambient down to 16\,K.

\begin{figure}
    \centering
    \includegraphics[width=0.95 \textwidth]{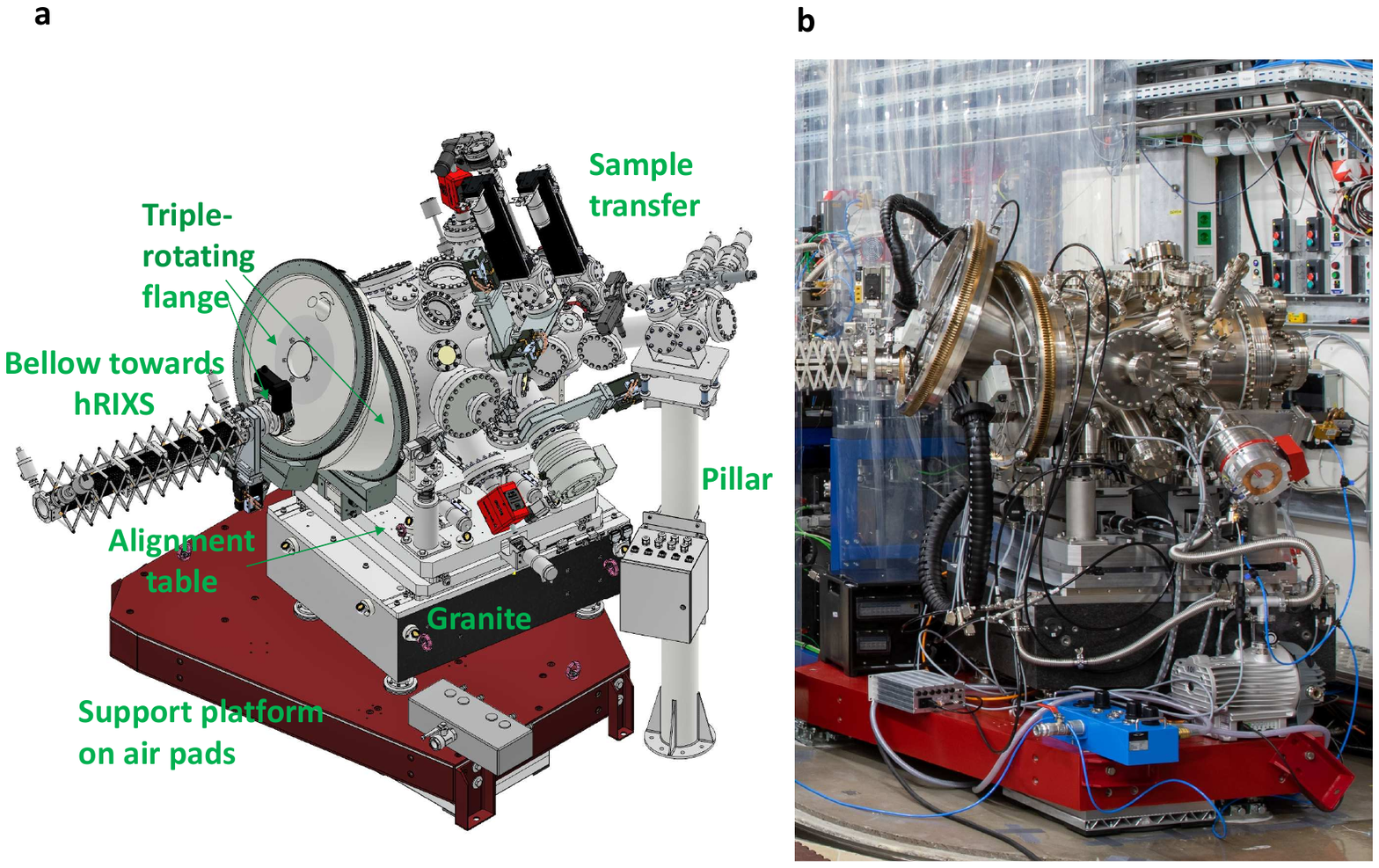}
    \caption{Overview of the XRD setup for solid sample environment: model (a) and photograph (b).}
    \label{fig:XRDchamber}
\end{figure}
    
Figure\,\ref{fig:XRDchamber} shows an overview of the setup from the outside. The core piece of the XRD vessel is the triple-rotating flange (TRF), which is the connecting flange to the hRIXS spectrometer. The TRF consists of three flanges, each one connected to a rotary feedthrough. Through a combined motion of the feedthroughs the connecting flange can move in the horizontal plane by about 90$^\circ$, which allows a continuous change of the scattering angle (see also Supplement). Besides to the hRIXS spectrometer, the flange can be alternatively connected to a 2D X-ray detector.

The XRD chamber is placed on a support platform with air-cushions (as is also the CHEM chamber), allowing to displace and place it back at the interaction point with precision below 100\,$\mu$m. The chamber is supported by a granite block which ensures good thermal and mechanical stability. A horizontal table above allows to align the chamber along and perpendicularly to the X-ray FEL beam. Vertical alignment is carried out by the inner mechanics placed on wedge shoes. The XRD setup is equipped with a sample transfer system.\\

\begin{table}
\caption{Motion ranges for XRD inner mechanics: detector circle (D.) and sample stage (S.) and the triple-rotating flange (TRF).}
\label{tab:XRDinnerMec}
\begin{tabular}{llll}
\hline \hline 
 Motion    &Range       &Repeatability    &Description        \\ 
\hline 
$2\theta$     &$\pm$180$^{\circ}$      &$<$ 0.002$^{\circ}$   &D. scattering angle     \\
$\theta$      &$\pm$180$^{\circ}$      &$<$ 0.002$^{\circ}$    &S. incidence angle      \\ 
$\chi$     &$\pm$30$^{\circ}$       &$<$ 0.002$^{\circ}$   &S. tilt      \\ 
$\phi$    &$\pm$90$^{\circ}$      &$<$ 0.1$^{\circ}$   &S. polar angle      \\ 
$tx$      &$\pm$5\,mm       &$<$ 5\,$\mu$m  &S. translation perp. to FEL \\ 
$tz$     &$\pm$5\,mm      &$<$ 5\,$\mu$m   &S. translation along FEL     \\ 
$ty$      &$\pm$5\,mm   &$<$ 5\,$\mu$m  &S. vertical translation \\ 
$2\Theta$      &60$^{\circ}$ - 147$^{\circ}$   &   &TRF scattering angle \\
\\
\end{tabular}
\end{table}

\begin{figure}
    \centering
    \includegraphics[width=0.95 \textwidth]{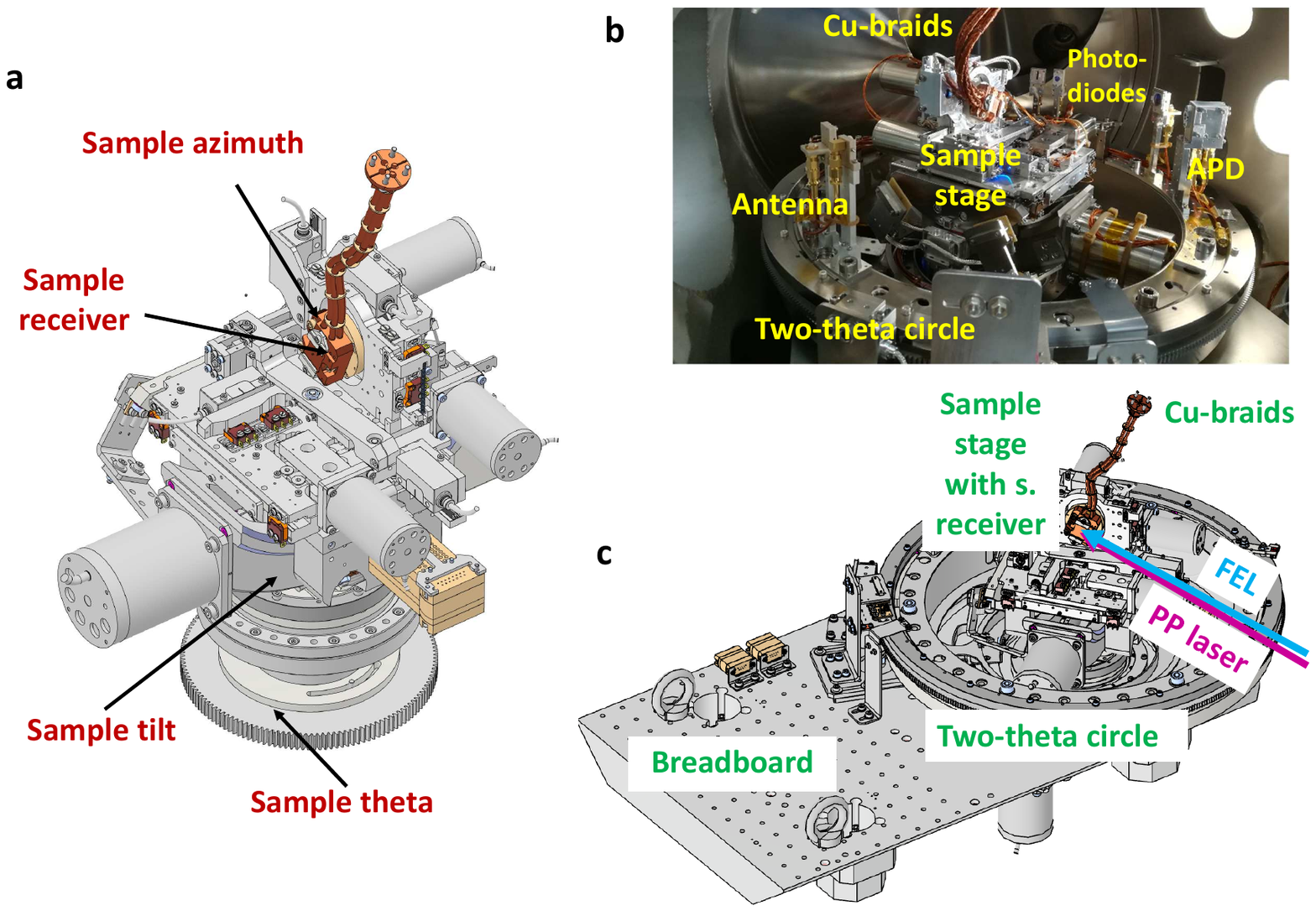}
    \caption{Inner mechanics of the XRD chamber: model revealing degrees of motion for the sample (a), photograph showing the diffractometer and the sample stage (b), model with overview of the inner mechanics (c).}
    \label{fig:XRDgonio}
\end{figure}

The inner mechanics consists of a diffractometer, a sample stage and a breadboard (see Figure\,\ref{fig:XRDgonio}). These parts form one entity that is supported directly by the wedge shoes and decoupled from the XRD vessel, making the entire structure internally stable. The diffractometer provides three degrees of motion, two for the sample stage (sample incidence angle $\theta$ and sample tilt $\chi$) and one for the detector circle (scattering angle $2\theta$). The sample stage is placed on top of the diffractometer and it provides another four degrees of motion for the sample (the three translations $tx$, $ty$, $tz$ and the polar angle $\phi$). Therefore, the diffractometer and the sample stage provide in total 6 degrees of motion for the sample. Figure\,\ref{fig:XRDgonio}a shows how the degrees of motion for the sample are staggered. Placing the translation stages above the diffractometer ensures that the rotation axis for $\theta$ and $\chi$ is always fixed and aligned with respect to $2\theta$. The polar angle $\phi$ is placed on top of the translation stages in order to keep the dimensions of the sample stage small. The movement range is listed in the table below (see also Supplement). All motors of the diffractometer and sample stage are in-vacuum stepper motors. The sample receiver is thermally disconnected from the sample stage and connected to a liquid-flow helium cryostat through Cu-braids, mounted on top of the XRD chamber. In order to keep the braids short while still allowing the full movement range of the the sample, the cryostat is placed on a manipulator that can follow the motion of the sample stage. In this configuration, temperatures below 20\,K can be easily obtained at the sample receiver (16\,K when the temperature at the cryostat is at 5\,K).

The detector circle allows mounting multiple photodiodes and APDs that are used for diagnostics and X-ray detection. Their location can be adopted to the experimental needs (in steps of 10$^\circ$). The detector circle also holds an "antenna" (SMA plug connected to a coax cable) that is used for timing diagnostics. The breadboard provides the possibility of mounting additional equipment inside the XRD chamber. In particular, it will be used to support a laser-incoupling mirror for experiments requiring short focal length for the pump-probe laser (e.g. mid-IR and THz wavelengths). For laser wavelengths where this is not required the laser-incoupling unit (LIN) of the SCS instrument is used (nearly-collinear incoupling geometry with 2\,m focal length for the XRD setup).

\section{First high-resolution RIXS spectra of hRIXS}

     \subsection{Settings and methods}

The hRIXS spectrometer was commissioned together with the CHEM endstation and its solid- and liquid-sample environment. To benchmark the performance static RIXS data was taken from well-studied sample systems. Measurements from solid samples were performed at room temperature and base pressure of $10^{-7}$\,mbar. We selected NiO and La$_2$CuO$_4$. The data demonstrates the high-resolution and high-throughput capabilities of the hRIXS instrumentation at SCS. Optimized spectrometer settings can be kept in the same position for weeks, without degradation of resolution, as was observed during operation. 
In order to demonstrate the feasibility of liquid-jet studies at hRIXS, we measured liquid water RIXS from a $40\,\mu$m cylindrical jet at a base pressure of $10^{-3}$\,mbar.

The available polarization of the XFEL beam was linar horizontal, i.e, lying in the scattering plane ($\pi$ polarization). All data was measured in the high energy-resolution mode, using the 3000\,l/mm grating of the spectrometer and the 150\,l/mm grating in the monochromator. The spectrometer was placed at a scattering angle of $2\Theta=125^\circ$. An integrating detector was used, PI-MTE3 2048\,x\,2048 CCD from Teledyne Princeton Instruments (pixel size of 15\,$\mu$m), mounted at $\gamma=25^{\circ}$ incidence angle (standard detector hRIXS geometry). The cooling temperature was set to $-50^\circ$\,C, the analog gain to high and digitization rate to 1\,MHz. The exposure time was several seconds (depending on photon energy and flux), such that single photon counting was possible. An optical filter with a membrane was installed between the CHEM vessel and hRIXS grating chamber, in order to block optical light and protect hRIXS vacuum from liquid-jet environment: 200\,nm Al film on 100\,nm Parylene C (from Luxel).

In order to initially find signal in the hRIXS spectrometer we used the diffraction peak from a multilayer sample. The final spectrometer alignment was the done on the specular reflection from a polished single-crystal sample (either NiO or CoO), which delivered the strongest non-resonant elastic signal with the $\pi$-polarized X-rays.  
 
    The measured RIXS spectrum appears as lines on the 2D detector. Due to non-corrected aberrations, those lines are not straight, but slightly parabolic, depending on the settings of the spectrometer. For each setting, we calibrate the aberrations by fitting a parabola to the most intense line (usually the elastic line). Using this calibrations we apply two methods to extract the actual spectrum, depending on the intensity of the signal:

    If the signal is weak such that individual photons can be seen, we fit those photons to a Gaussian peak, which reveals the position of the photon even more accurate than one pixel \cite{amorese2019enhanced,kummer2017rixstoolbox}. For strong signals we employ a simple integration strategy: For each pixel we calculate its central energy following the aberration correction, then summing the intensity of pixels into one-pixel-wide bins, where the intensity is split between the two closest bins, proportionate to how close the central energy is to either bin.

\begin{figure}
    \centering
    \includegraphics[width=0.7 \textwidth]{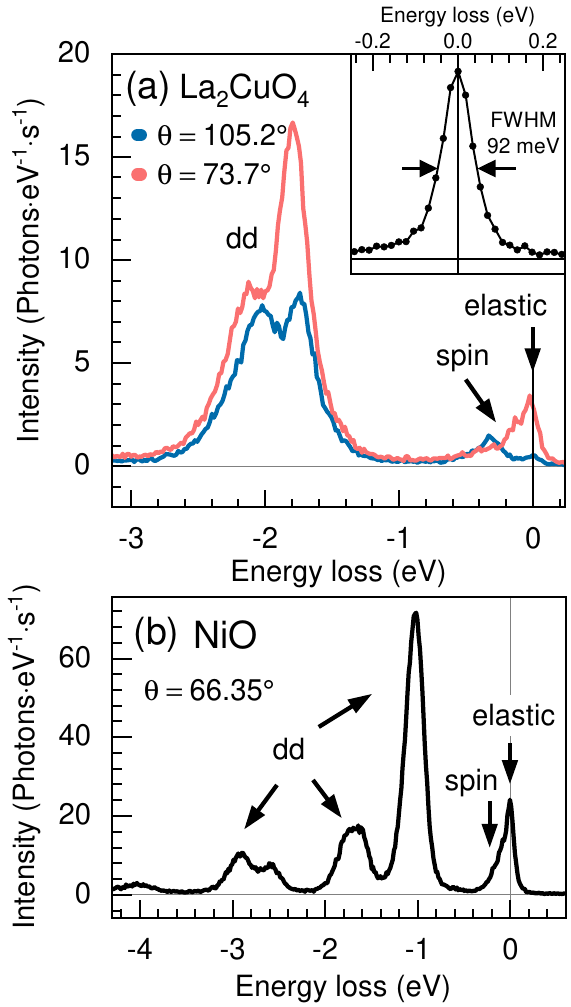}
    \label{fig_LCOspectra}
    \caption{(a) RIXS spectra of thin-film \ce{La2CuO4} at Cu L$_3$ edge measured at two different incident angles. The surface normal of the sample is parallel to the crystalographic $c$-direction. The inset shows elastic line near specular conditions to estimate the combined resolution. (b) RIXS spectrum of bulk-crystal NiO at Ni L$_3$ edge. The surface normal is the crystallographic $a$-direction. The elastic, magnetic and dd excitations are indicated in the spectra.}    
\end{figure}

    The data analysis is done on-line, i.e. after each image taken the analysis is automatically performed. The data is stored in the XFEL data acquisition system, and can later be analysed off-line. We have developed a toolbox \cite{scstoolbox} containing the relevant analysis procedures. It is based on the Python xarray package \cite{hoyer2017xarray}, and is usually used within jupyter notebooks \cite{kluyver2016jupyter}.

     \subsection{Experimental data from solid samples}

Figure\,\ref{fig_LCOspectra}(a) shows two Cu L$_3$ RIXS spectra of a $\sim$48\,nm thick La$_2$CuO$_4$ film grown on LaSrAlO$_4$ and a Ni L$_3$ edge RIXS spectrum of a single crystal NiO sample. The measured combined resolution of 93\,meV\,FWHM at Cu L$_3$ edge (80\,meV\,FWHM at Ni L$_3$ edge) was obtained with $\sim$8\,$\mu$m $\times$ 200\,$\mu$m focus on the sample. The samples were at room temperature and the measurements were done with an intra-train repetition rate of 113\,kHz 
and an attenuation factor of the X-ray beam of 30$\%$ that gives $\sim$0.22\,$\mu$J/pulse at Cu L$_3$ edge and $\sim$0.17\,$\mu$J/pulse at Ni L$_3$ edge as measured by the X-ray gas monitor. In these conditions the samples did not show evident alteration due to radiation damage, as it could be judged from the absorption and RIXS spectral shape. The count rate is comparable to that of equivalent spectra measured at synchrotron beamlines \cite{braicovich2010magnetic,betto2021multiple,ghiringhelli2009observation,betto2017NiO}. The repetition rate is compatible with $\sim$100\,kHz pump laser repetition rate, and can be increased by a factor of 10 (see Supplementary Materials for NiO data measured at 1.13\,MHz repetition rate). The limiting factor would be the pump laser pulse energy at high repetition rates (particularly when using the OPA) and the sample damaging when increasing the average power irradiated through both the XFEL and the pump laser. The spectra correspond to different scattering geometries, resulting in different projection of the electrical-field vectors for the incident and exit X-ray beams on the crystallographic axes, as well as different momentum transfer along the Cu-O plane. 
As for \ce{La2CuO4}, at $\theta\simeq74^\circ$ the in-plane component for the $dd$ excitation peaks is more pronounced and the spin excitations have single magnon and bi-magnon contributions at comparable intensity. For $\theta\simeq105^\circ$, where the momentum transfer is larger, the single magnon is better resolved, 
while its scattering cross section is bigger than that of the bi-magnon \cite{ament2009theoretical,braicovich2010momentum,bisogni2012bimagnon}. 
As for NiO (Figure\,\ref{fig_LCOspectra}(b)), the $dd$ features are all well resolved, while the spin excitations are at the limit of the experimental resolution. These spectra show that RIXS measured on prototypical strongly correlated materials and excited with ultra-short and intense XFEL pulses preserves their shape and can be interpreted in the same way as those measured at synchrotrons.

     \subsection{Experimental data from liquid samples}

\begin{figure}
    \centering
    \includegraphics[width=0.75 \textwidth]{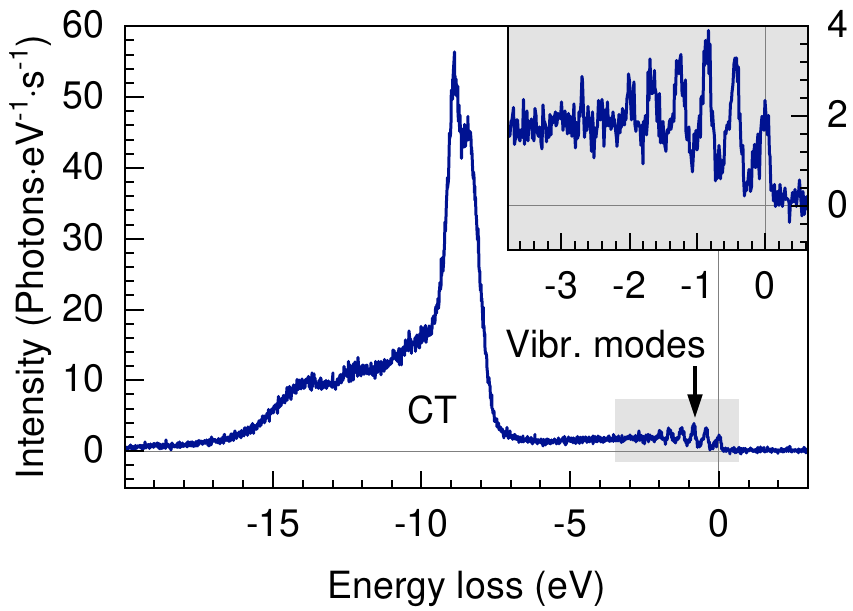}
    \caption{RIXS spectrum of liquid water obtained during commissioning at room temperature with the 1000\,l/mm grating. The inset shows a close-up of the vibrational progression.}
    \label{fig:WaterSpectrum}
\end{figure}

The spectrum of liquid water from liquid jet measured at 535\,eV corresponding to the maximum of the O K-edge pre-edge feature is shown in Figure\,\ref{fig:WaterSpectrum}. The spectrum exhibits the electronic excitations previously reported between 8 and 18\,eV energy loss for liquid-cell measurements \cite{weinhardt2012, harada2013selective, vaz2019probing, pietzsch2022cuts} and liquid jet \cite{yin2015ionic, lange2010high}. The inset shows a close-up of the vibrational progression. This spectrum was measured at 1.13\,MHz repetition rate, 450\,pulses/train, 100\% transmission and $100\,\mu$m beamline exit slit (FEL output power of $500\,\mu$W). The polarization of the X-rays was linear horizontal (leading to a reduced cross section for vibrational excitations, compared to linear vertical polarization) and the spectrometer was placed at $2\Theta=125^\circ$. The diameter of the liquid jet was $40\,\mu$m and the flow rate was set to 1.2\,ml/min. The resulting spectrum was constructed from 20\,minutes of acquisition. The pressure in the experimental chamber was at $\sim$4.5$\cdot{}10^{-3}$\,mbar.

     \subsection{Performance of hRIXS at SCS}

Table\,\ref{tab:hRIXS_meas_performance} displays an overview of the measured performance. Combined resolving power above 10000 is reached in the photon energy range of 500\, to 1000\,eV with the current high-resolution monochromator grating. At 1.1\,MHz repetition rate, high-resolution RIXS from transition-metal monoxides (e.g. NiO) can be obtained within few minutes. The data throughput increases by about factor 5 when working with 1000\,l/mm spectrometer grating (compared to 3000\,l/mm). For experiments where modest resolving power of 3500 is sufficient, the flux can be potentially increased by another factor 6 when using the high-throughput monochromator grating. This is particularly interesting for liquid-jet experiments.\\ 

\begin{table}
\caption{Achieved performance when using current high-resolution monochromator grating. hRIXS angular acceptance of about 47\,mrad$^2$ (with horizontal detector size of 30.7\,mm). Photon flux at 1.1\,MHz repetition rate and 400\,pulses/train with $100\%$ transmission, $50\,\mu$m beamline exit slit and spot size of $12\,\mu$m x $100\,\mu$m\,(v x h). Machine at 11.5\,GeV. Whith high-transmission monochromator grating the photon flux on the sample increases by about factor 6, while the resolving power decreases to about 3500.}
\label{tab:hRIXS_meas_performance}
\begin{tabular}{llll}      
\hline \hline 
 Photon   &Photon flux    &hRIXS    &Combined re-       \\ energy\,(eV)     &at sample\,(ph/s)    &grating  &solving power          \\ 
\hline
850      &$1.0\cdot10^{-13}$    &3000\,l/mm\,\,\,\,\,\,      &10600     \\
930      &$1.3\cdot10^{-13}$   &3000\,l/mm    &10100      \\ 
530      &$1.6\cdot10^{-12}$   &1000\,l/mm    &10600        \\
850      &as above    &1000\,l/mm      &\,\,\,7080       \\ \\
\end{tabular}
\end{table}

The present energy resolution is already sufficient to study crystal-field and spin excitations in cuprates and in other 3d transition metal oxides. However, at the moment, the combined resolving power is limited by the beamline monochromator. When the high-resolution grating will be installed, the combined band width will drop to $\sim45$\,meV at the Cu L$_3$ edge and comparatively even better at lower energies, where the edges of Ni, Co, Fe, Mn, Cr, V and Ti can be found. This means that not only orbital, charge-transfer and spin excitations, but also phonons and charge excitations \cite{rossi2019experimental,arpaia2019dynamical} can be studied in a pump-probe fashion, exploiting the high-resolution RIXS at SCS.

As the presented data demonstrates, RIXS and XES signals of liquids  
do not deviate from their spectra collected at soft X-ray synchrotron sources. It suggests that radiation-damage processes are not observed with the presented hRIXS experiments (i.e. happen after the detection of the scattered soft X-ray photons). This is potentially due to the use of the ultrashort soft X-ray pulse widths (100\,fs and smaller), the 1\,$\mu$m focus, which allows a 3-times exchange of the sample in the soft X-ray focus at given jet speed, thus leading to the MHz-rapid sample exchange protocol used for the liquid jet.
For experiments in solution chemistry it follows - given the described hRIXS sensitivity and the MHz repetition frequency - that it is possible to investigate the metal edge of i.e. metal organic compounds down to 10\,$\mu$M concentration regimes. (Our estimates assume single-shot detection at one selected X-ray energy: $10^{11}$\,ph/pulse, 0.5\,$\mu$m focus, typical soft X-ray penetration depth of 200\,nm at the jet’s surface, typical soft X-ray cross sections: ca.\,$\approx 700$\,Mbarns $= 7 \times\,10^{-20}$\, m$^2 = 7$\,{\AA}$^2$.) The same holds true also for edges of soft elements of 1st periodicity in the table of chemical elements, whose K-shell electrons are excited with soft X-rays. Hence, the sensitivity of a chemistry experiment has been increased by up to 4 orders of magnitudes compared to conventional synchrotron experiments. This has a major impact on the type of photochemistry studied. In optical spectroscopy it is a well-known fact that the chemistry of a system deviates heavily when molecules are studied at 10-100\,mM concentrations compared to $\mu$M concentration regimes. ''Golden rules'' allow only reliable studies in $\mu$M concentration regimes, when molecular (non-periodic) properties of chemical matter and chemical reactions are investigated.

\section{Summary and Outlook}

The combination of the high repetition rate European XFEL with the brilliance driven, photon hungry technique of RIXS turns out to be a perfect match as seen from the swift data acquisition in the combined high energy resolution and femtosecond time information. Thus, excited state detection and ultrafast dynamics is now routinely accessible with temporal resolution between 30\,fs and 100\,fs and an energy resolution of some tens of meV. In consequence, the full potential of dynamic RIXS 
is pushed towards the fundamental information content given by the transform limit in energy and time. This performance is achieved through the optical design of a VLS spherical-grating geometry, which has been optimized to balance energy range and resolving power, as well as throughput. In particular, the energy range from 500\,eV to 1000\,eV with a maximum resolution of 50\,meV at the
oxygen K-edge has been achieved. In the future an additional grating can be mounted for further optimized performance on specific spectral regions or extending the photon energy range into the tender X-rays. The strengths and uniqueness of the chosen optical and mechanical design of the hRIXS spectrometer is that resolution is currently set by the energy resolution of the SASE3 beamline grating and the pixel size of the X-ray detectors. We have intentionally built into the optical and mechanical design the ability to harvest future improved performance parameters on the soft X-ray SASE 3 beamline and detector technologies. Foreseeable are the developments of large high-resolution beamline gratings, detectors with smaller pixel size and advanced timing modes. In particular, differential measurements between ultrafast stimulus (i.e. laser) "on" vs. "off" at a rate within the European XFEL bunch trains will minimize the averaging influence of inter- and intra-bunch train jitter and further improve the statistics. Here, the rapid evolution of pixelated and fast read-out detectors is going to benefit hRIXS. In consequence, enhanced pump on-off contrast at highest duty cycle opens the path towards dilute active sites and potentially novel excitations with lower cross-sections, such as dressed states of the ground state and pump-induced transient states. The modular design of the sample environments in combination with the confocal set-up of beamline focus and hRIXS spectrometer source point, allows to exchange optimized sample environments without loosing the alignment of the infrastructure. Currently, a dedicated UHV solid-state sample environment and liquid-phase chemistry environment are in operation. However, potentially other complementary environments can be envisaged in the future. The availability of variable linear and circular polarization through the SASE3 Apple-X afterburner at the soft-X-ray branch of European XFEL is going to give additional contrast mechanisms such as RIXS-XMCD (X-ray Magnetic Circular Dichroism), where the dynamic response of magnetically ordered and chiral materials and molecules is going to benefit.


\section{Acknowledgements}
We thank all members of the Heisenberg RIXS (hRIXS) international user consortium for initiating, designing and funding the hRIXS project. We thank Nick Brookes, Kurt Kummer and Flora Yakhou-Harris of the ESRF for helping with the preliminary CCD detector and reference RIXS spectra. GG acknowledges invaluable discussions with Nick Brookes and Lucio Braicovich on the design of RIXS spectrometers. We thank Mark Dean, Xi He and Ivan Bozovic for providing the La$_2$CuO$_4$ sample. We acknowledge all participants of user-assisted commissioning. Funding by the Initiative and Networking Fund (IVF) of the President of the Helmholtz Association in the framework of Helmholtz International Beamlines at European X-FEL is acknowledged. Funding by the ERC-ADG-2014 Advanced Investigator Grant No. 669531 EDAX under the Horizon 2020 EU Framework Program for Research and Innovation is acknowledged. Funding by the Academy of Finland research infrastructures grant 303914 is acknowledged. Funding by the Italian Ministry of Research (MIUR): PIK project ``POLARIX'' is acknowledged. Funding by the Deutsche Forschungsgemeinschaft (DFG) (217133147/SFB 1073 project C02) is acknowledged.

\referencelist

\end{document}